\documentclass[twocolumn,pre,floatfix,superscriptaddress]{revtex4}

\newcommand{\beq}{\begin{equation}}
\newcommand{\eeq}{\end{equation}}
\newcommand{\bea}{\begin{eqnarray}}
\newcommand{\eea}{\end{eqnarray}}

\newcommand{\be}{\begin{equation}}
\newcommand{\ee}{\end{equation}}
\newcommand{\ba}{\begin{eqnarray}}
\newcommand{\ea}{\end{eqnarray}}
\newcommand{\rd}{\mathrm{d}}            
\newcommand{\re}{\mathrm{e}}            
\newcommand{\ri}{\mathrm{i}}

\newcommand{\te}[1]{\mbox{\boldmath $ #1 $}}

\newcommand{\bx}{\te{x}}
\newcommand{\by}{\te{y}}
\newcommand{\bn}{\te{n}}

\newcommand{\bu}{\te{u}}
\newcommand{\bv}{\te{v}}

\newcommand{\bbr}{\te{r}}

\newcommand{\bz}{\te{z}}

\newcommand{\bF}{\te{F}}
\newcommand{\bT}{\te{T}}

\newcommand{\bR}{\te{R}}
\newcommand{\bV}{\te{V}}

\usepackage{graphicx}
\usepackage{times}
\usepackage{color}
\usepackage{amsmath}
\usepackage{epstopdf}
\begin{document}

\title{Flexible helical yarn swimmers}
\author{A.~P. Zakharov, A.~M. Leshansky and L.~M. Pismen}
\email{andreiz@technion.ac.il}
\affiliation{Department of Chemical Engineering, Technion -- Israel Institute of Technology, Haifa 32000, Israel}

\begin{abstract}
We investigate the motion of  a flexible Stokesean flagellar swimmer realised as a yarn made of two intertwined elastomer fibres, one active, that can reversibly change its length in response to a local excitation causing transition to the nematic state or swelling, and the other one, a passive isotropic elastomer with identical mechanical properties. A propagating chemical wave may provide an excitation mechanism ensuring a constant length of the excited region. Generally, the swimmer moves along a helical trajectory, and the propagation and rotation velocity are very sensitive to the ratio of the excited region to the pitch of the yarn, as well as to the size of a carried load. External excitation by a moving beam is less effective, unless the direction of the beam is adjusted to rotation of the swimmer.

\end{abstract}
 
 \maketitle

\section{Introduction}

Stokesean swimmers have been a lively field of research starting from pioneering work of Taylor \cite{Taylor} and Lighthill \cite{Lighthill}, driven both by the interest to ways of locomotion of microorganisms \cite{Lauga,Elgeti} and by design of biomimetic artificial microswimmers that could be remotely actuated, navigated and delivered to a specific location \emph{in vivo} (see \cite{wang,gaowang} for state-of-the-art reviews). Most artificial swimmers are driven either chemically or magnetically. The first category includes Janus particles with catalytically active or reactive surfaces fuelled by interaction with the external medium \cite{paxton,howse}. Driving by a rotating magnetic field enables contact-free and fuel-free propulsion  without chemical modification of the environment. Magnetically driven propulsion requires non-trivial coupling between translational and rotational motions of the object. Most often, magnetic swimmers have a chiral geometry, and are structured as either artificial helical flagella \cite{N2,GF} or propellers of regular \cite{GF,ML14} or random \cite{random} shape. It is also possible to fabricate polarisable superparamagnetic (not possessing remanent magnetisation) helical propellers \cite{PetersIEEE13}.

All above mentioned modes of propulsion are quite different from natural locomotion based on internal chemical driving and employing complex mechanisms of transmission of chemical to mechanical energy in living tissues. Another characteristic of natural locomotion is a flexible body change of the swimmer. Although flexible magnetic swimmers have been described and tested, their flexibility is not essential for locomotion but is caused by feedback of the viscous stress on the elastic swimmer \cite{LaugaSM,Coq} or elastic instabilities in the magnetic field \cite{Py}. Biohibrid inherently flexible swimmers have been constructed by using DNA strands to bind magnetic particles \cite{dna} and employing selectively cultured contractile cells to deform elastomer filaments \cite{cell} or sheets shaped as a jellyfish \cite{jfish}. Some theoretical recipes for chemical swimmers imitating basic features of natural locomotion, though not its intrinsic mechanism, include reaction-diffusion processes in compliant materials \cite{Balazs} and stimulation by gel swelling \cite{Alexeev}. Swimming due to local phase transitions in nematic elastomers was demonstrated by an example of a nematoelastic sheet swimming into the dark \cite{swim}. Very recently, propulsion of a nematoelastic ``worm" was implemented by exciting travelling waves of radial expansion and longitudinal contraction \cite{fisher}.

\begin{figure}[t]
\includegraphics[width=.4\textwidth]{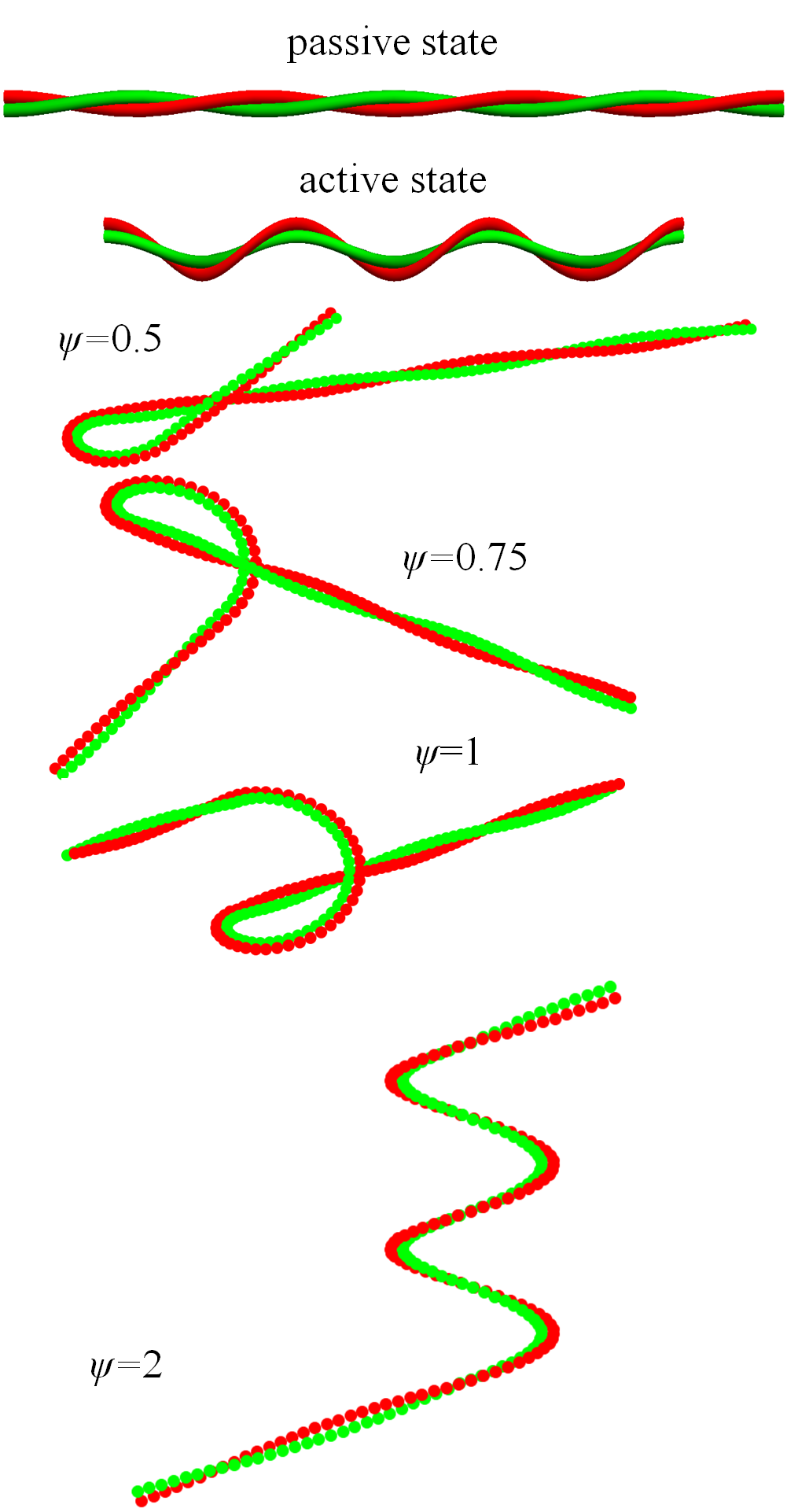}
 \caption{Above: a twisted yarn in the passive state and after nematic transition or swelling. Other pictures show the shapes or a 3-pitch yarn (discretised by spheres for the hydrodynamic computation) with a middle active segment comprising a fraction $\psi$ of the pitch, as indicated. }
 \label{f:fiber}
\end{figure}

In this communication, we propose and investigate a flexible Stokesean swimmer in the form of a \emph{twisted yarn} made of two intertwined elastomer fibres, one active, that can reversibly change its length in response to a local excitation, and the other one, a passive isotropic elastomer with identical mechanical properties, glued along a contact line (Fig.~\ref{f:fiber}). This simple configuration featured recently as a nematoelastic crawler \cite{azlp} reshaping due to a propagating actuating wave inducing reversible isotropic-nematic transition (INT) that extends locally the axially polarised active fibre. Alternatively, elongation may be caused by local swelling of an elastic hydrogel  \cite{Balazs}. The two cases differ only by a change of the fibre radius $r$, which grows in the same proportion as the length in the hydrogel but shrinks to preserve the volume in nematic elastomers. This, however, only weakly affects locomotion of a slender fibre.  The most natural driving mechanism for either swelling or INT is an oscillatory chemical reaction \cite{Balazs,gel10}.

\section{Computation algorithm}
\subsection{Reshaping due to reversible phase transition}

When the active fibre elongates locally by a factor $\lambda$, the curvature develops along the normal vector directed along the line connecting the centres of both fibres, which rotates with the original pitch. As a result, the contact line, being rectilinear before extension, turns into a helix with the curvature radius $R$ verifying the equation $\lambda/ (R+ r)=1/(R-r)$, or $R \approx  2r/\epsilon$ at  $\lambda-1 \equiv \epsilon \sim r \ll R$. Besides the intrinsic curvature $\overline{\kappa}=1/R$ dependent on the local nematic elongation or swelling, the yarn has a constant intrinsic torsion $\overline\tau = 1/\ell$, where $\ell$ is the pitch measured along the contact line.

The instantaneous shape of the yarn is computed by minimising the elastic energy density\cite{AP} (scaled by the shear modulus and the yarn area $2\pi r^2$) dependent on the differences between observed strain $u$, curvature $\kappa$ and torsion $\tau$ and their intrinsic values $\overline u = \lambda, \, \overline{\kappa}, \,\overline\tau$:
\begin{equation}
\mathcal{E} = \frac 12  (u-\overline{u})^2
 +  \frac {C_b}{2} (\kappa -\overline{\kappa})^2 +
 \frac {C_t}{2} (\tau-\overline{\tau})^2.
 \label{eq:Elast1}
\end{equation}
The bending rigidity factor equal to the bending moment relative to the contact line $4\int_{-r}^r (x+r)^2 \sqrt{r^2-x^2} dx=\frac 52 \pi r^4$ divided by the yarn area is $C_b=\frac 54 r^2$. The torsional rigidity factor $C_t=\frac 12 r^2$ is the same as for a single fibre with the circular cross-section \cite{AP}.

We will further assume both INT (or swelling) and elastic reshaping to be fast. We will refer to a  nematic or swollen segments as ``active" and isotropic or unswollen ones as ``passive". Neglecting narrow transition zones at the border between both kinds of segments, the energy can be set to zero by choosing the shape of the contact line satisfying $ u=\overline{u}, \,\kappa = \overline{\kappa}, \,\tau=\overline{\tau}$.  Starting from the passive state, the active region propagates with a constant speed along the yarn. Some intermediate shapes are shown in Fig.~\ref{f:fiber}. In the particular case when the length of the active domain equals to the pitch of the yarn or its multiple, the two passive domains at its opposite ends are parallel; otherwise, they form a finite angle.

\subsection{Particle-based hydrodynamic computation}\label{alg}

The solution of the hydrodynamic problem is based on the particle-based approach \cite{KK91,filippov00}, which has been previously applied for modelling the various low-Reynolds-number swimmers, e.g., a rotating helical filament \cite{lesha09}, Purcell's toroidal swimmer \cite{LK08}, and an undulating flexible filament \cite{BKSL12}. Both fibres of the twisted yarn are discretised by chains of spheres as shown in Fig.~\ref{f:fiber}. If a load is added, it is modelled  by a sphere of a larger radius $\rho$. Translation and rotation velocities $\mathbf{V}_i (t)=\{\mathbf{u}_i(t),\, \omega_i(t)\}$ of each sphere are defined in the comoving (but not corotating) coordinate frame at each time step $t$ by displacements due to advancement of the excitation wave. The choice of a coordinate frame is not essential, since the overall displacement and rotation are eventually determined by the hydrodynamic force and torque balance.

The procedure described in the Appendix is applied at each time step to determine the generalised force vector \cite{Brenner} $\mathcal{F}(t)=\{\mathbf{F},\mathbf{T}\}$ comprising the force \textbf{F} and torque \textbf{T} acting on the swimmer. The three components of the velocity \textbf{U} of the swimmer and three angles ${\Omega}$ (components of the Euler vector) characterising the rotation velocities of the object at each time step are found from the requirement of force- and torque-free locomotion. The unknown velocities are calculated using the grand resistance matrix $\mathcal{K}(t)$ that relates the vector $\mathcal{F}(t)$ given by Eq.~(\ref{totFT}) and the unknown generalised velocity vector \cite{Brenner} ${\mathcal U}=\{\mathbf{U},\, {\Omega}\}$:
\begin{eqnarray}
\label{resmatrix}
 \mathcal{U}(t)= -(\mu \mathcal{K})^{-1}\mathcal{F}(t),
\end{eqnarray}
where $\mu$ is dynamic viscosity.
The components of the resistance matrix $\mathcal{K}(t)$ are determined, exploiting the linearity of the Stokes equation, by computing the corresponding forces and torques due to translation/rotation of the yarn along/about the axes of the laboratory frame.

The local velocities of the surface and hence, the vector ${\cal U}$ are determined in a comoving coordinate frame that is rotated with respect to the laboratory framework at each step of the computation. The actual trajectory of the swimmer in the \emph{laboratory} frame is found by rotating the displacement at each computation step $\mathbf{U}(t)$ with the help of the instantaneous rotation matrix $ \mathbf{R}({\Upsilon}(t))$, where ${\Upsilon}(t) =\int_{0}^{t} {\Omega}(t') dt' $ is the accumulated rotation at the step $t$, and integrating the rotated propulsion velocity:
 \begin{eqnarray}
\label{path}
\mathbf{X}( t ) =\int_{0}^{t}  \mathbf{R}({\Upsilon}( t' )) \cdot \mathbf{U}( t' ) dt'.
\end{eqnarray}
\section{Results and discussion}
\subsection{One-cycle and multicycle trajectories}

The computation algorithm described in the preceding Section was applied to trace displacement and rotation of the swimmer during an excitation wave traversing its length and inducing local transition to the nematic state or swelling. In the beginning and the end of each excitation cycle, the contact line of the two strands of the yarn is rectilinear but is both displaced and rotated.

The total rotation per cycle can be characterised by comparing the orientations of both straight lines, and complementing this by \emph{intrinsic} rotation about the swimmer's axis, which does not change the orientation of the contact line but only shifts the phase of the pitch and thereby, the orientation of the normal and binormal in the fundamental plane at the ``head' of the yarn (defined at the point of the entry of the excitation wave). The rotation angles were computed, in accordance to the Tait-Bryan convention, in the following specified sequence: first, executing intrinsic rotation by an angle $\phi$ about the tangent vector \textbf{t}, then, rotating by an angle $\theta$ about the binormal \textbf{b}, and finally, by an angle $\gamma$ about the normal \textbf{n} in the fundamental plane.

All three rotation angles affect multicycle trajectories, since displacements at the next cycle should be rotated accordingly. Given the displacement per cycle and the rotation matrix $\mathbf{R}(\phi,\theta,\gamma)$, the average motion of the yarn through a number of excitation cycles can be determined with minimal additional computations. The displacement vector at any $n$th cycle is computed as $\mathbf{X}_n =\mathbf{R}(\phi,\theta,\gamma) \cdot \mathbf{X}_{n-1}$. The average displacement is obtained by iterating this formula. Generally, the multicycle paths are quasiperiodic helices, unless all rotation angles are commensurate with $\pi$.

\subsection{Motion driven by chemical wave}

\begin{figure}[h]
 \centering
\includegraphics[width=.4\textwidth]{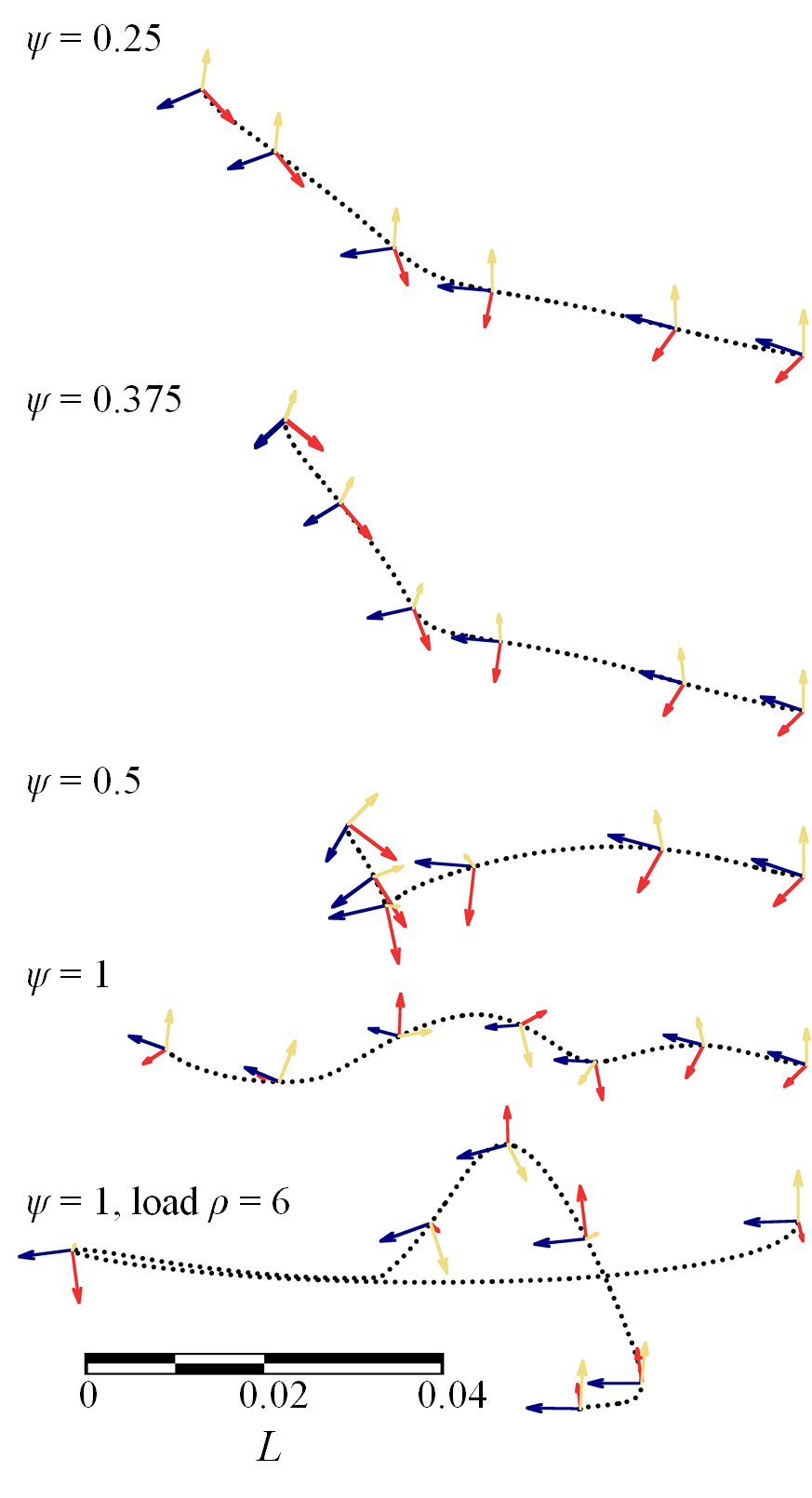}
 \caption{Examples of  paths during a single cycle at different values of  the ratio $\psi$ with the common scale bar. The arrows show instantaneous directions of the tangent (blue, or the darkest), normal (red, or dark gray), and binormal (yellow, or light gray) vectors at the head of the yarn. The excitation wave moves to the right, and the swimmer moves to the left.}
 \label{f:OnePathsPoints}
\end{figure}

We have investigated in detail the case of excitation due to an oscillatory chemical reaction. In this case, the length of the active segment measured along the contact line remains constant as the wave propagates. The motion is strongly dependent on the ratio $\psi$ of the excited region to the pitch of the yarn $\ell$, as well as on the size of a load. Some trajectories of the geometric centre of the yarn during a single cycle are shown in Fig.~\ref{f:OnePathsPoints}. Take note of the effect of an added load leading to a convoluted trajectory.

\begin{figure*}
 \centering
\begin{tabular}{cc}
(a) & (b) \\
\includegraphics[width=.4\textwidth]{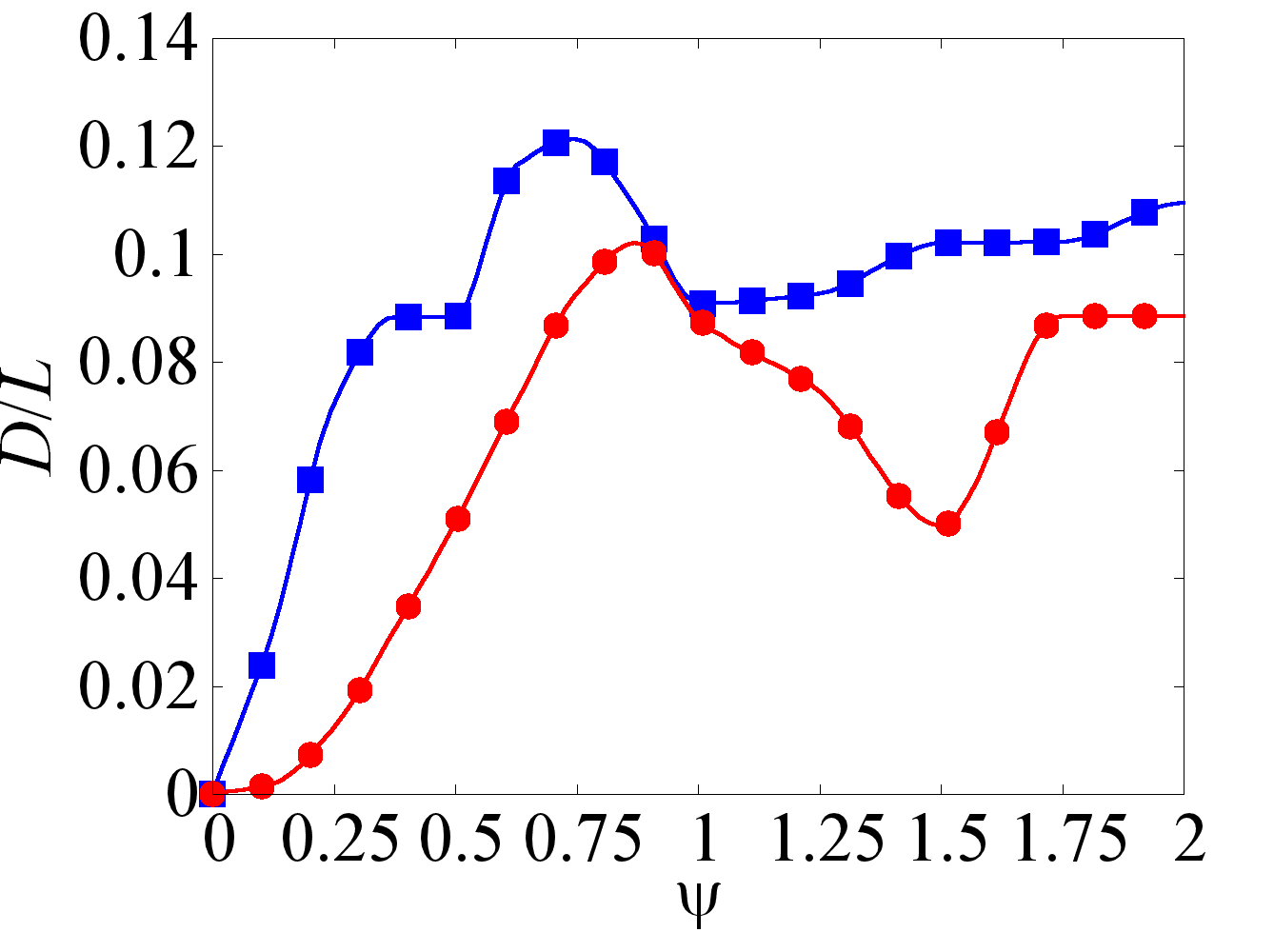} &
\includegraphics[width=.4\textwidth]{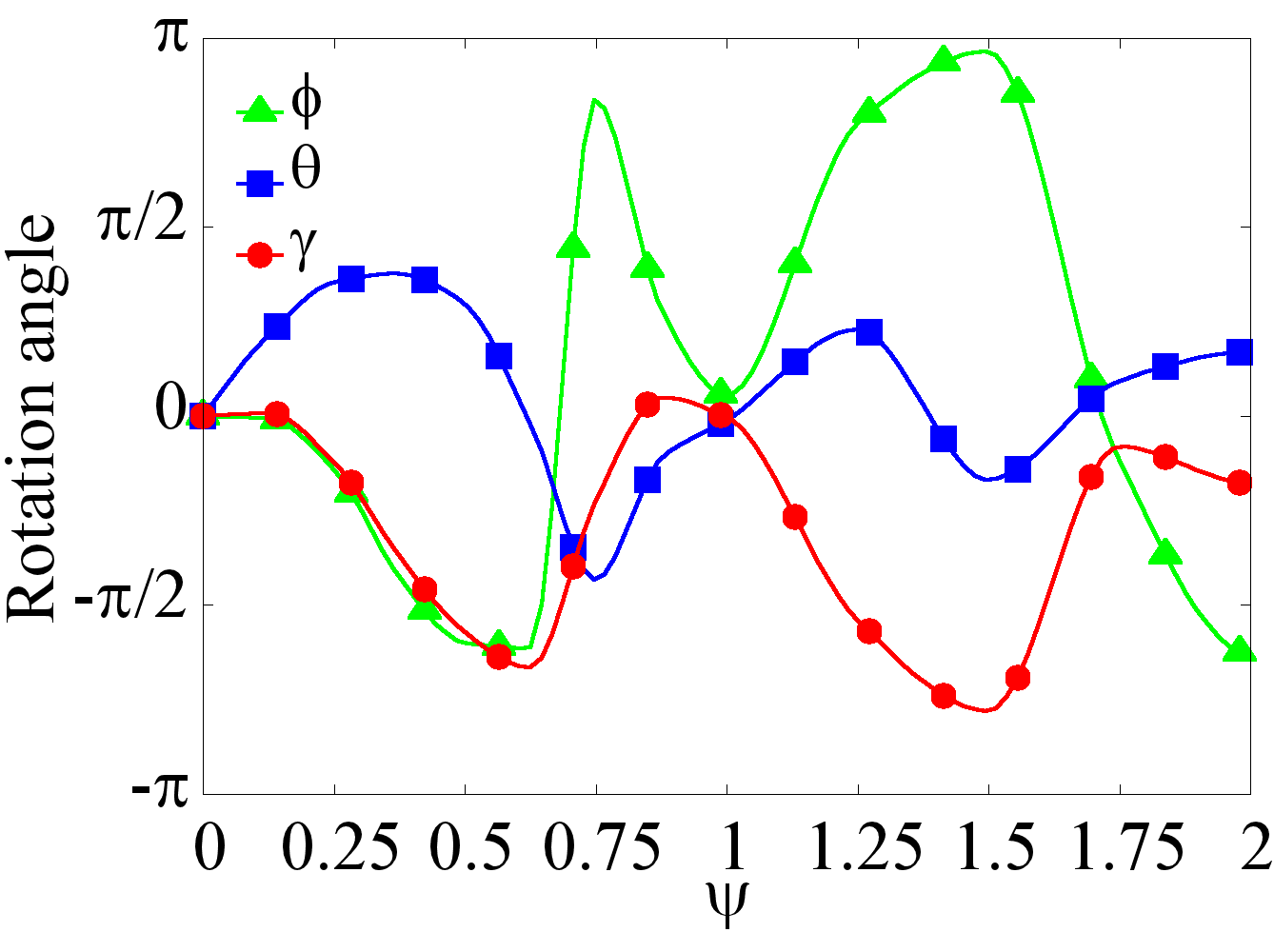} \\
(c) & (d)\\
\includegraphics[width=.4\textwidth]{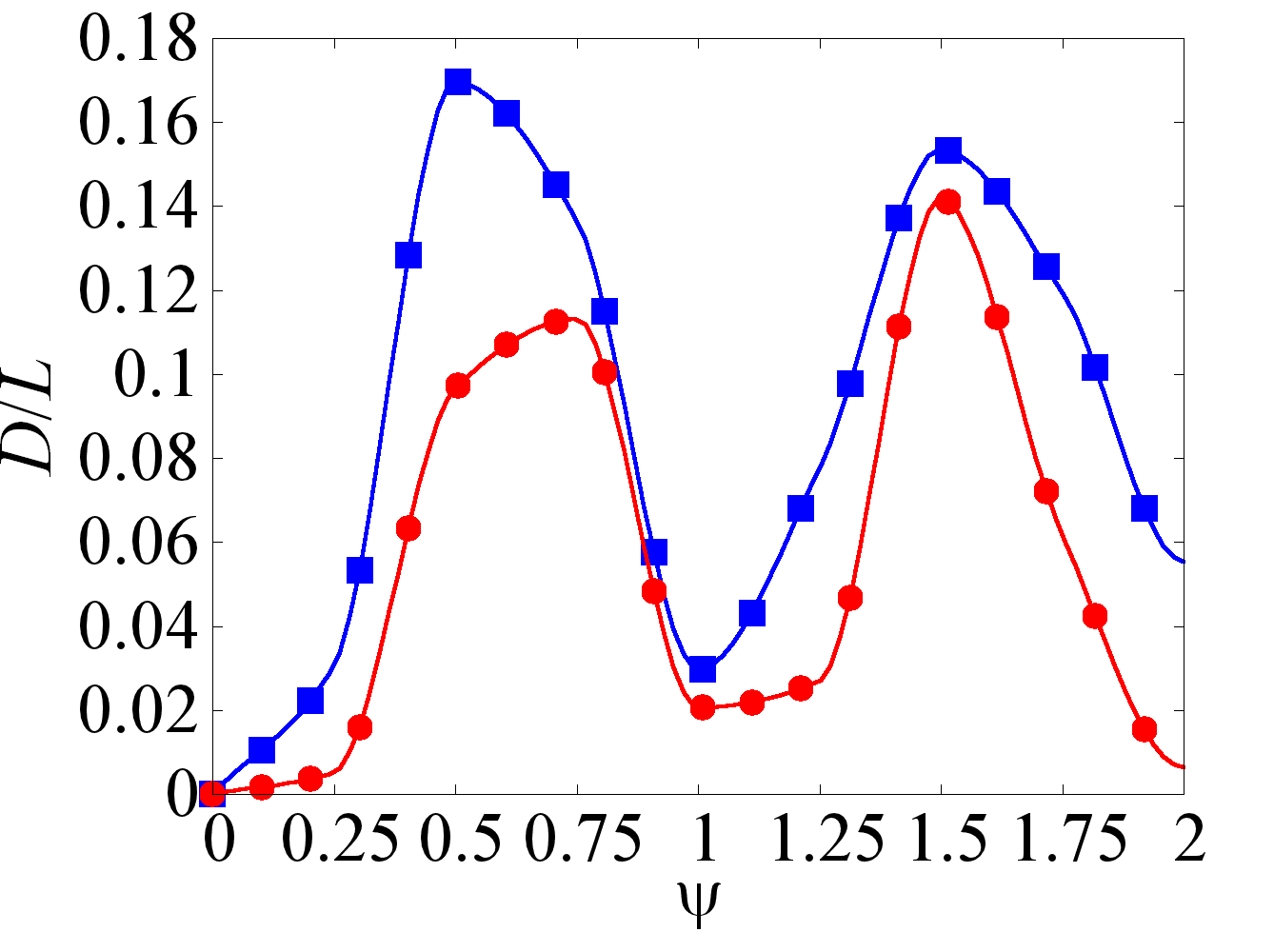} &
\includegraphics[width=.4\textwidth]{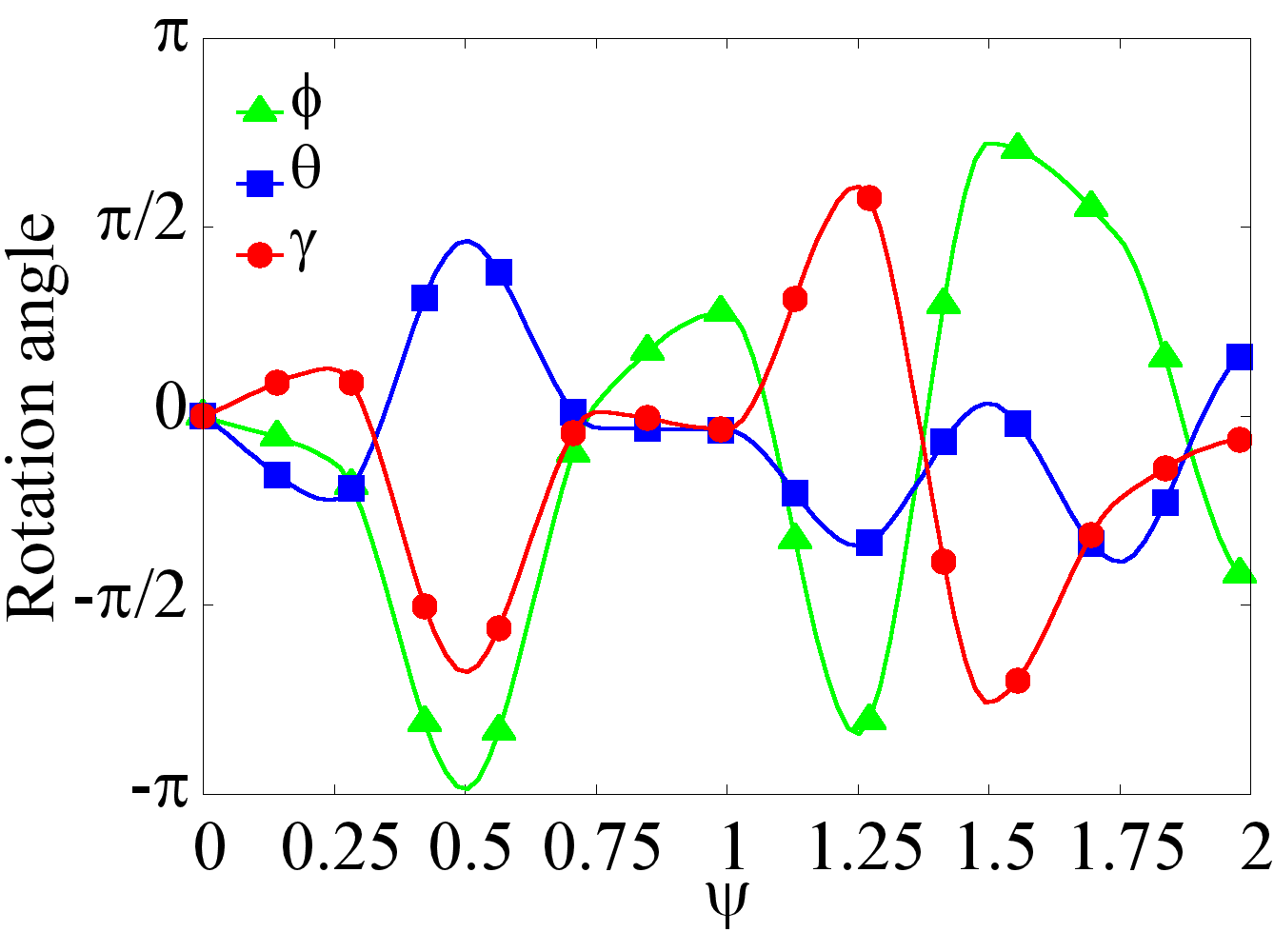} \\
(e) & (f)\\
\includegraphics[width=.4\textwidth]{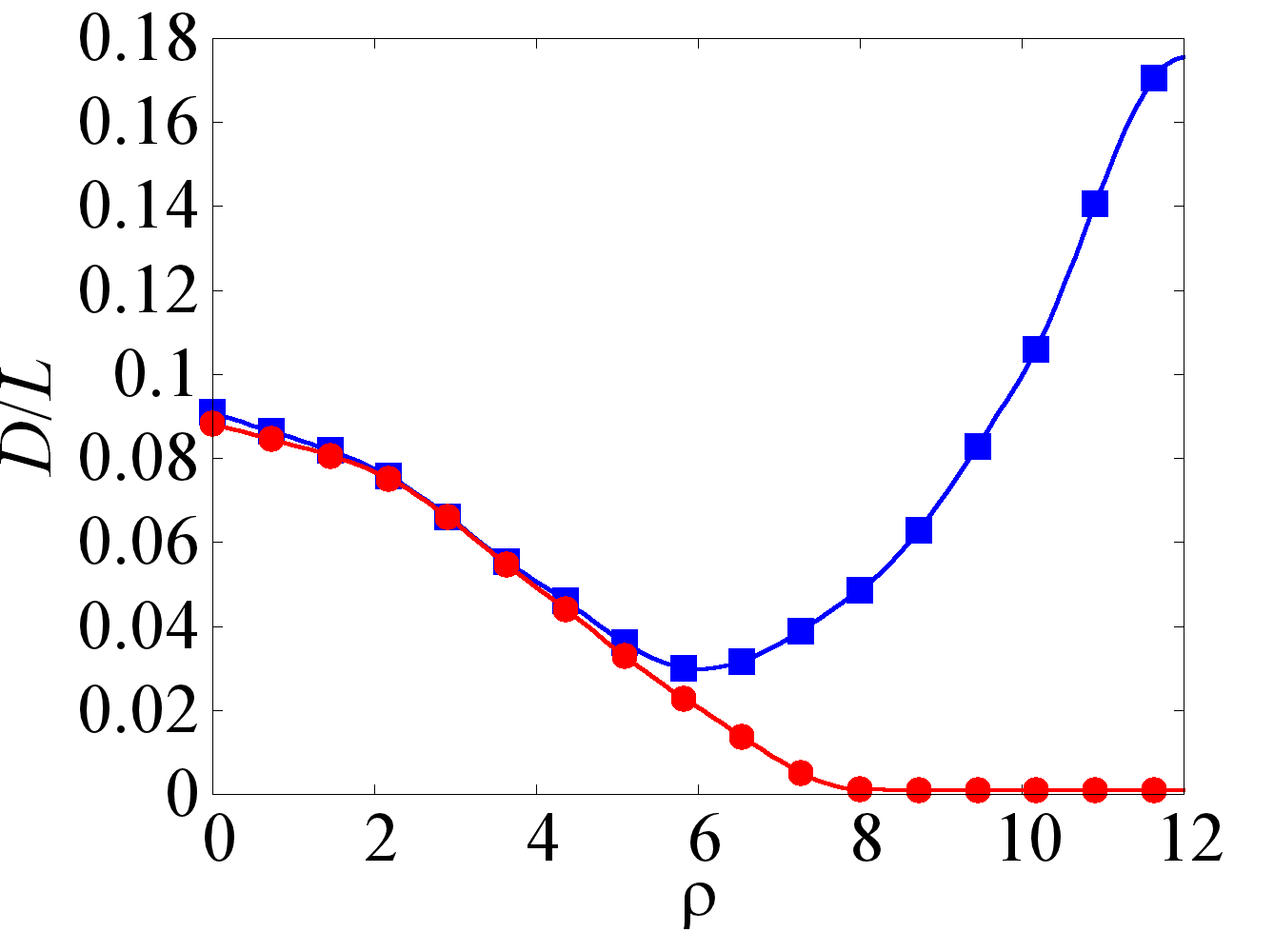} &
\includegraphics[width=.4\textwidth]{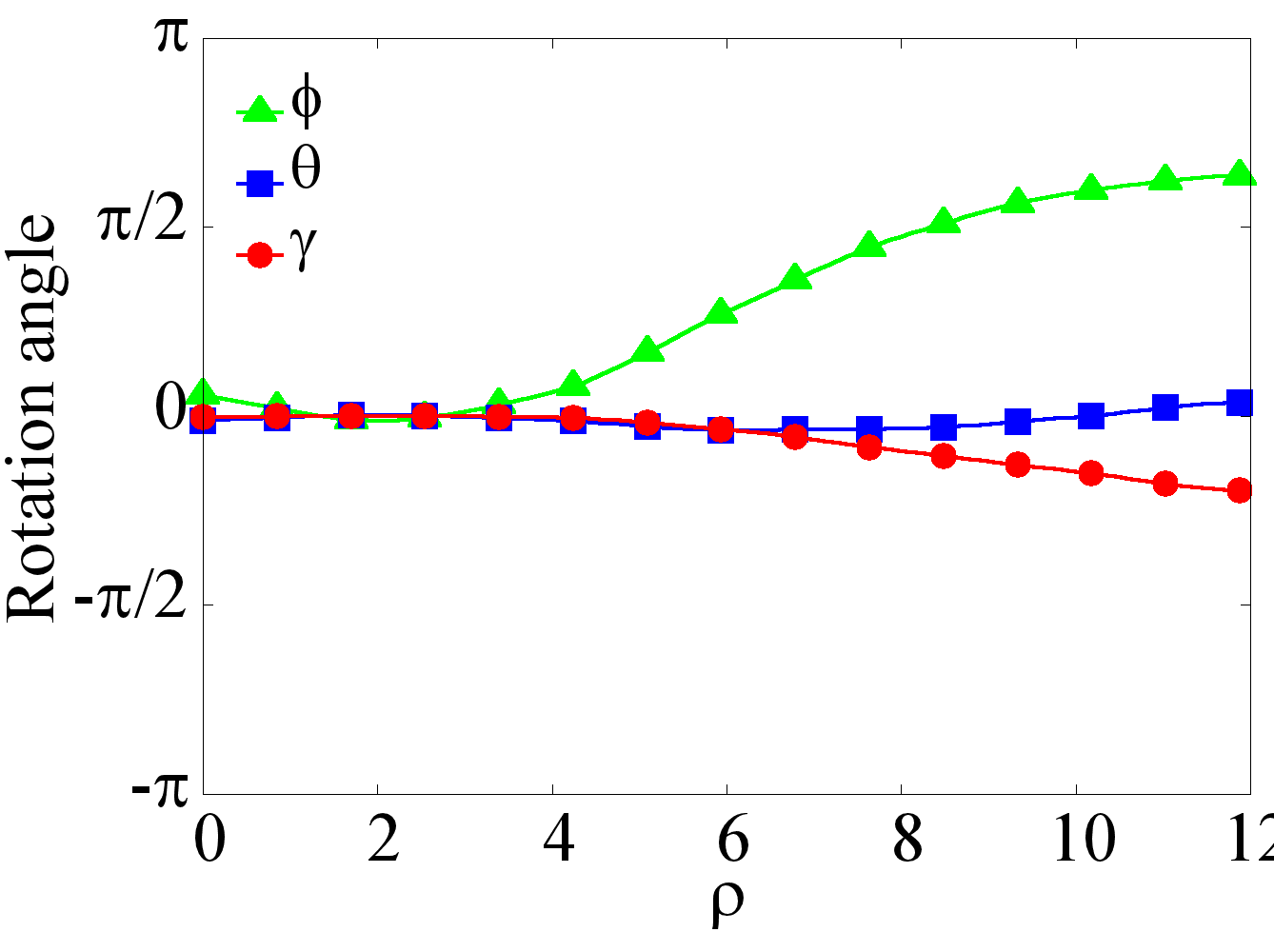}\\
(g) & (h)\\
\includegraphics[width=.4\textwidth]{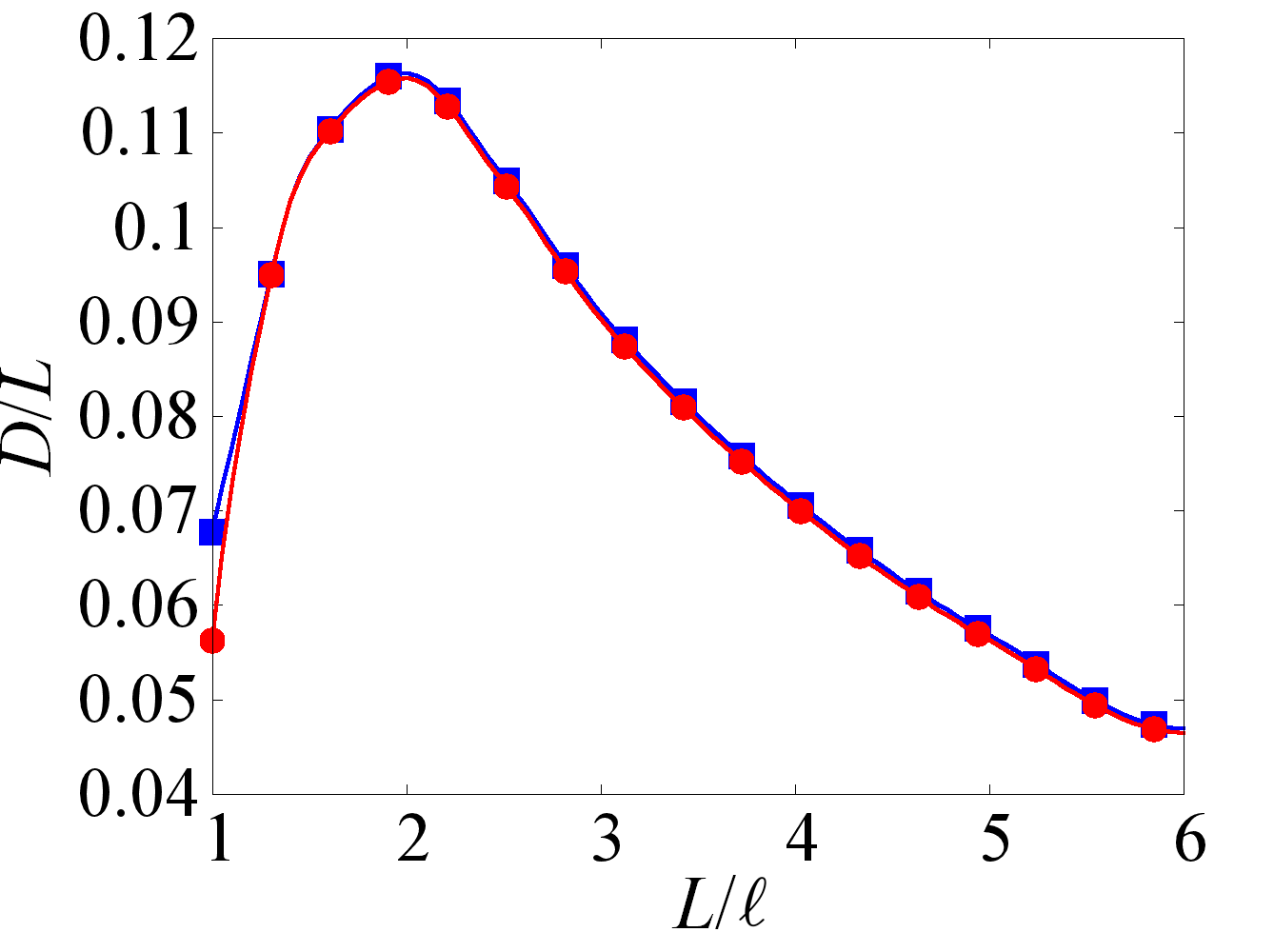} &
\includegraphics[width=.4\textwidth]{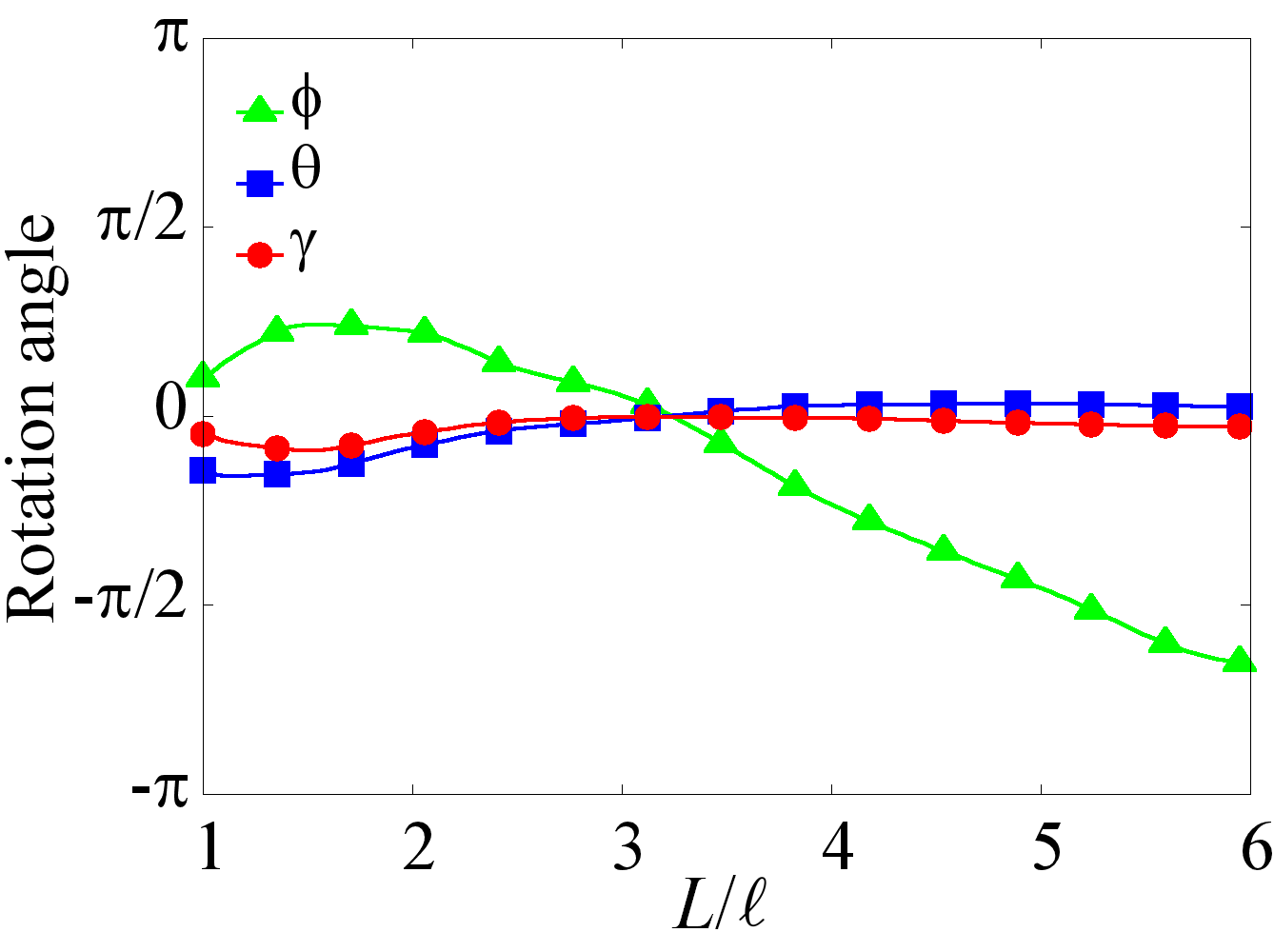}
\end{tabular}
  \caption{Two upper rows: the dependence of the ratio of advancement to length ratio $D/L$ (left panels) and rotation angles (right panels) on the ratio $\phi$ of the excited region to the pitch for a yarn with the length $L=3\ell$  (a,b) and the same yarn carrying a load (c,d). Two lower rows: the dependences of $D/L$ and rotation angles on the load size $\rho$ (e,f) and the number of pitches $L/\ell$ (g,h) at $\psi=1$ and no load. The upper lines (marked by squares) in the left panels show the displacement per a single passage of the excitation wave, and the lower lines (marked by circles), the average displacement along the stationary direction of motion after a number of cycles.}
 \label{f:displ}
\end{figure*}

The dependences of the advancement $D=|\mathbf{X}|$, where \textbf{X} is the translation vector of the geometric centre, on $\psi$ for a yarn with the length $L=3\ell$ and the same yarn carrying a load -- a sphere with the diameter six times larger than that of a fibre -- are shown in Fig.~\ref{f:displ}(a,c). The upper (marked by squares) and lower (marked by circles) curves in these Figures show, respectively,  the displacement per single passage of the excitation wave and the average displacement along the stationary direction of motion after a number of cycles. The dependence is rather irregular but swimmers carrying a load move fastest when the length of the active region is either one-half or three-halves of the pitch. Figs.~\ref{f:displ}(e,g) show the dependence of the displacement on the load size $\rho$ and the number of pitches $L/\ell$. Remarkably, the displacement per cycle \emph{grows} at larger loads but average displacement decreases monotonically, dropping down to negligible values at $\rho>8$.

The difference between the displacement per cycle and average displacement is due to helical motion caused by persistent rotations. The right panels of Fig.~\ref{f:displ} show the three components of the Euler vector that characterises the change of orientation of the contact line of the yarn. The dependence on the ratio $\psi$ in Fig.~\ref{f:displ}(b,d) is also irregular, while rotation at $\psi=1$ shown in Fig.~\ref{f:displ}(f)  becomes significant upon increasing the size of the load.

\begin{figure}[t]
 \centering
 \includegraphics[width=.4\textwidth]{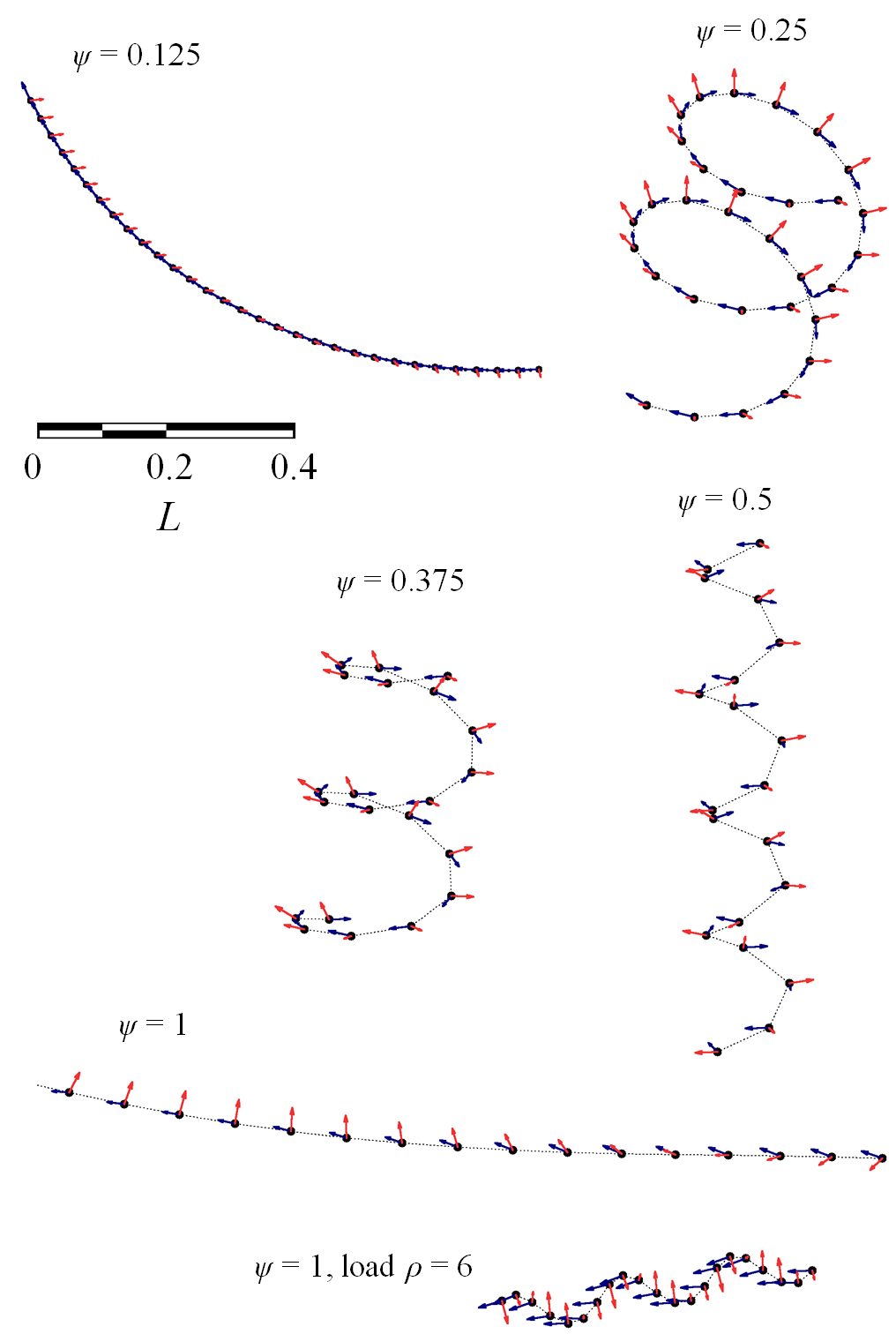}
 \caption{Examples of helical paths showing the common scale bar. Circles indicate positions at beginning/end of each cycle, and arrows show the directions of the tangent and normal vectors at these moments.}
 \label{f:MultiPathsPoints}
\end{figure}

 One can see in Fig.~\ref{f:MultiPathsPoints} that the trajectories are very sensitive to both the value of the ratio $\psi$ and the load. In particular, a large difference between the displacement per cycle and average displacement for heavy loads is due to a highly convoluted trajectory of a loaded yarn. The displacement per cycle and average displacement do not differ significantly only when both $\theta$ and $\gamma$ are close to zero, independently of $\phi$. This is seen, in particular, in Fig.~\ref{f:displ}(g,h) drawn for the case $\psi=1$ with no load. In this case, the multi-cycle trajectory is almost rectilinear. As these two angles increase, the width of the helical trajectory and, as a result, the difference between the displacement per cycle and average displacement rapidly increase, as seen, for example, for yarns with large loads in Fig.~\ref{f:displ}(e,f).

\subsection{Motion driven by guiding beam}

\begin{figure}[t]
 \centering
 \includegraphics[width=.4\textwidth]{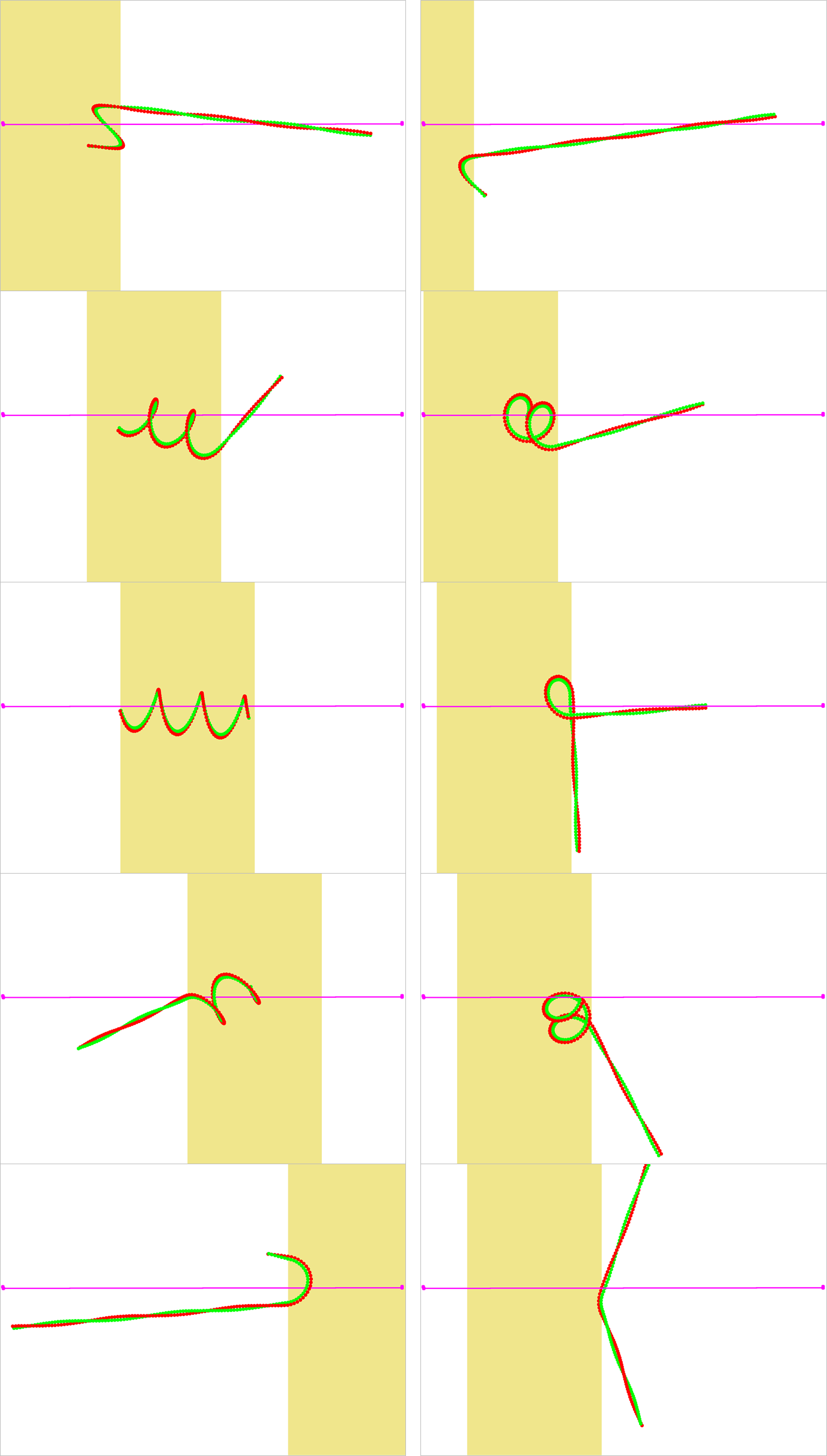}
 \caption{Snapshots of the yarn excited by a constantly oriented moving beam, taken during the first (left) and second (right) cycle. The horizontal line shows the initial orientation of the yarn, and shading, the position of the beam moving from the left to the right.}
 \label{f:ThickBeam}
\end{figure}

If the yarn is excited by an external ''beam'', the length of the excited region is not constant but increases in the bent segments; it also increases when the yarn is oriented at an acute angle to the beam. This may cause trapping of the swimmer in the beam, as illustrated by snapshots in Fig.~\ref{f:ThickBeam} taken for the case of a wide beam covering one pitch in the initial passive state. During the first cycle, shown in the left panel of this Figure, the excited segment increases as it shifts to the right in accordance to the beam motion, but nothing extraordinary happens before the swimmer straightens again and rotates. At a certain stage during the second cycle, however, as the excited segment further increases due a changed orientation, the swimmer becomes totally trapped within the beam, and rapidly changes its shape, as seen in the right panel of  Fig.~\ref{f:ThickBeam}. Advancement becomes ineffective under these conditions but is improved if the orientation of the beam changes to keep it normal to the axis of the swimmer as it turns after the cycle has been completed. In this case, the absolute value of the displacement is the same for each cycle, equal to  $D = 0.0795\, L$, and the average multicycle displacement is $0.0566\, L$. 

Such an entrapment is avoided when the beam is narrower than the width if the helical yarn in the bent state. In the case of a narrow beam with the length of 1/8 pitch, the displacement of a three-pitch swimmer per one cycle is $0.02308 L$, while the average multicycle displacement is $0.02234\, L$ with the maximum rotation angle $0.0651\pi$. If  this relatively small rotation is compensated by rotating the beam accordingly, the average displacement rises to $0.02297 L$, meaning that the overall motion is almost rectilinear. Although the last number is smaller than the above value for a wide beam with adjustable orientation, the propulsion efficiency, estimated by assuming the energy input to be proportional to the beam length, is much higher than for the wide beam.

\subsection{Conclusion}

The composite yarn is perhaps the simplest realisation of a biomimetic swimmer. Although it looks superficially similar to a bacterial flagella, its mode of operation is completely different, being based on continuous torque-free reshaping rather than flagellum rotation driven by torque applied by the molecular motors at the bacterial head. In some sense, this propulsion mechanism is similar to locomotion of Spirochetes lacking external flagella but driven by rotation of the internalised flagella yielding similar time-periodic reshaping \cite{CG02}. Indeed, any differential rotation varying along the length of the swimmer would cause propulsion with efficiency dependent on the rotation protocol.

Powering the swimmer by oscillatory chemical reaction (possibly fueled by reactants carried in the attached load) appears to be the most efficient mode of autonomous motion, but thorough coordination between the period of the chemical wave and the pitch of the yarn is essential for efficient multicycle propulsion. Guiding by an external beam is rather problematic, due to possible trapping, and can be made more efficient if the beam orientation is adjusted by tracing the position and orientation of the swimmer.

\emph{Acknowledgement} This research is supported by Israel Science Foundation  (grant 669/14).  AML acknowledges the support by the German--Israeli Foundation (grant I-1255-303.10/2014).

\appendix \section{Hydrodynamic computation algorithm}
The general solution for the velocity and pressure fields around a collection of $N$ spherical particles of radii $\mathrm{a}_i$, can be written as superpositions
\begin{equation}
\mathbf{v}=\sum_{i=1}^{N}{\bv_i}\:,\quad p=\sum_{i=1}^{N} P_i
\end{equation}
where the solution for the velocity $\bv_i$ outside a single $i$th sphere has the form of Lamb's general solution of Stokes equations in terms of solid spherical
harmonics \cite{KK91},
\begin{eqnarray}
& &\bv_i=\bv_i'+\frac{1}{2\mu} \bbr_i P_i=\sum_{n=1}^{\infty}\left[{\nabla \times \left(\bbr_i \chi^{i}_{-\left(n+1\right)}\right)+\nabla \Phi^i_{-\left(n+1\right)}}\right. \nonumber \\
&&-\left. \frac{\left(n-2\right)}{2\mu n\left(2n-1\right)}r_i^2 \nabla p^i_{-\left(n+1\right)}+\frac{\left(n+1\right)}{\mu n \left(2n-1\right)}\bbr_i
p^i_{-\left(n+1\right)}\right]
\nonumber \\
\end{eqnarray}
Here $\bbr_i$ is the radius vector with origin at the centre of the $i$th sphere, $r_i=|\bbr_i|$; $p^i_{-(n+1)}$, $\chi^i_{-(n+1)}$, and $\Phi^i_{-(n+1)}$  are combinations of solid harmonics arising, respectively, from the solution of the associated homogeneous equations $\nabla^2 P_i'=0$, $\nabla\cdot \bv_i=0$, and $\nabla^2 \bv_i'=0$:
\beq
\left\{\Phi^{i}_{-\left(n+1\right)},\frac{1}{\mu}p^i_{-\left(n+1\right)}, \chi^i_{-\left(n+1\right)} \right\}=\sum_{m=-n}^n \left\{ a_{mn}^i, b_{mn}^i, c_{mn}^i\right\}\: u_{mn}^{i-}\, ,\nonumber \\
\eeq
with $u^{i-}_{mn}$ being decaying solid spherical harmonics of the order $-(n+1)$ centred at the origin of the $i$th sphere:
\beq
u_{mn}^{i-}=\frac{1}{r_i^{n+1}}P_n^m\left(\cos{\theta_i}\right)\re^{\ri m \phi_i},\label{eq:harmonics}
\eeq
where $P_n^m$ is the associated Legendre function. For $n=1$, the solutions $\{\Phi^i_{-2},\,\frac{1}{\mu} p^i_{-2},\,\chi^i_{-2}\}$ correspond, respectively, to a stresslet, stokelet and rotlet centred at the $i$th sphere \cite{KK91}.

The no-slip boundary conditions, $\bv=\bV_i$, where $\bV_i$ is the local velocity of the surface of $i$th particle following from the solution of elastic problem are used to determine the unknown coefficients $a_{mn}^i$, $b_{mn}^i$ and $c_{mn}^i$. The feedback of flow on the yarn shape is assumed to be negligible due to weakness of the hydrodynamic, relative to the elastic, forces.

An elegant way of computing the coefficients was proposed by Filippov \cite{filippov00}. The boundary conditions are first transformed to the Lamb's form by applying the operators $\bbr_i\cdot$, $-r_i\nabla\cdot$, and $\bbr_i\cdot\nabla\times$ to both sides of the no-slip boundary condition , followed by the direct origin-to-origin transformation of spherical harmonics centred at different spheres, yielding an infinite system of linear equations for the coefficients $X_{mn}^i$, $Y_{mn}^i$ and $Z_{mn}^i$ in the expansions of ${\bbr_i\over r_i} \cdot \bu_i$, $-r_i\nabla\cdot \bu_i$ and $\bbr_i\cdot\nabla\times \bu_i$ in surface harmonics:
\begin{eqnarray}
\label{fil1}
&&a_i^{n+1} X_{mn}^i = -(n+1)a_{mn}^i+\frac{(n+1)}{2\:(2n-1)}b_{mn}^i \nonumber \\
&& +a_i^{2n+1}\sum_{j=1}^{N}{\sum_{l=1}^{\infty}{\sum_{k=-l}^l{\left(D_{klmn}^{ij}a_{kl}^j+E_{klmn}^{ij}b_{kl}^j+F_{klmn}^{ij}c_{kl}^j\right)}}},
\nonumber \\
\end{eqnarray}
\begin{eqnarray}
\label{fil2}
&&a_i^n Y_{mn}^i={1\over a_i^2}(n + 1)(n+2)a_{mn}^i-\frac{n(n+1)}{2\left(2n-1\right)}b_{mn}^i \nonumber \\
&&\quad  +\sum_{j=1}^{N}{\sum_{l=1}^{\infty}{\sum_{k=-l}^l{\left(G_{klmn}^{ij}a_{kl}^j+H_{klmn}^{ij}b_{kl}^j+L_{klmn}^{ij}c_{kl}^j\right)}}},
\nonumber \\
\end{eqnarray}
\begin{eqnarray}
\label{fil3}
&& a_i^{n+1} Z_{mn}^i=n\left(n+1\right)c_{mn}^i \nonumber \\
&& +a_i^{2n+2} \sum_{j=1}^{N}{\sum_{l=1}^{\infty}{\sum_{k=-l}^l{\left(M_{klmn}^{ij}b_{kl}^j+N_{klmn}^{ij}c_{kl}^j\right)}}}.
\nonumber \\
\end{eqnarray}

Here $a_i$ stand for sphere radii; the coefficients $D_{mnkl}^{ij},\: E_{mnkl}^{ij},\: F_{mnkl}^{ij},\:K_{mnkl}^{ij},\: L_{mnkl}^{ij},\:M_{mnkl}^{ij}$ and $N_{mnkl}^{ij}$ are given in the appendix of \cite{filippov00} in terms of the transformation coefficient $C^{ij}_{klmn}$:
\[
C^{ij}_{klmn}=(-1)^{m+n}\frac{(l+n-k+m)!}{(l-k)!(m+n)!} u^{j-}_{(k-m)(l+n)}(R_{ij},\theta_{ij},\varphi_{ij})\:,
\]
where $R_{ij},\theta_{ij},\varphi_{ij}$ are the spherical coordinates of vector $\bR_{ij}$ connecting the centres of $j$th and $i$th spheres, and $u^{j-}_{(k-m)(l+n)}$ are the decaying solid spherical harmonics defined in (\ref{eq:harmonics}). According to the definition of spherical harmonics, the coefficients $C_{klmn}$ are set to zero if $|k|>l$ or if $|m|>n$.

When the particle surface velocity corresponds to the rigid body motion, $\bV_i=\bu_i+\te{\omega}_i\times\bbr_i$, the right hand sides of Eqs.~(\ref{fil1}) -- (\ref{fil3}) can be written as \cite{filippov00}:
\begin{eqnarray}
\label{X11}
X_{1n}^i&=&\frac{1}{2}\left(V_{ix}^0-\ri V_{iy}^0\right)\delta_{n}^1, \quad
X_{0n}^i=V_{iz}^0\delta_{n}^1,\nonumber\\
X_{-1n}^i&=&-\left(V_{ix}^0+\ri V_{iy}^0\right)\delta_{n}^1, 
\end{eqnarray}
\begin{eqnarray}
\label{Z11}
Z_{1n}^i&=&\left(\omega_{ix}^0-\ri \omega_{iy}^0\right)\delta_{n}^1,\quad
Z_{0n}^i=2\omega_{iz}^0\delta_{n}^1,\nonumber\\
Z_{-1n}^i&=&-2\left(\omega_{ix}^0+\ri \omega_{iy}^0\right)\delta_{n}^1,
\end{eqnarray}
where $\{\bu_i, \mbox{\boldmath$\omega$}_i\}$ are, respectively, the translation and rotation velocities of $i$th sphere, and $\delta^k_n$ is the Kronecker's delta. The coefficients $X_{mn}$, $Z_{mn}$ for $|n|>1$ vanish identically, and so do all $Y_{mn}$.

The viscous  force $\bF_i=\int_{\partial S_i} \te{\sigma\cdot}\bn\:\rd S$ and torque $\bT_i=\int_{\partial S_i} \te{r}_i\times (\te{\sigma\cdot}\bn)\:\rd S$ exerted on $i$th sphere about its centre can be expressed in terms of the expansion coefficients for $n=1$,

\begin{eqnarray}
\label{filforce}
\bF_i=-4\pi\mu
\left[\left(b_{11}^i-\frac{1}{2}b_{-11}^i\right)\hat{\bx}+\ri\left(b_{11}^i+\frac{1}{2}b_{-11}^i\right)\hat{\by}+b_{01}^i
\hat{\bz}\right] \nonumber 
\end{eqnarray}
\begin{eqnarray}
\end{eqnarray}
\begin{eqnarray}
\label{filtorque}
\bT_i=-8\pi\mu
\left[\left(c_{11}^i-\frac{1}{2}c_{-11}^i\right)\hat{\bx}+\ri\left(c_{11}^i+\frac{1}{2}c_{-11}^i\right)\hat{\by}+c_{01}^i
\hat{\bz}\right] \nonumber 
\end{eqnarray}
\begin{eqnarray}
\end{eqnarray}

Thus, when velocities of the spheres are prescribed, the forces and torques exerted on any sphere can be found by solving $3\:N\times L\times(L+2)$ equations for the expansion coefficients $\{a_{mn}^i, b_{mn}^i, c_{mn}^i\}$, obtained by truncating the system (\ref{fil1}-\ref{fil3}) after $l=L$ terms and solving it together with Eqs.~(\ref{filforce}) -- (\ref{filtorque}). Alternatively, the forces and torques can be prescribed and velocities computed, or a mixed problem can be formulated with some velocities and forces/torques prescribed.

The instantaneous net force $\bF$ and torque $\bT$ exerted on the object built of $N$ spheres is computed by straightforward superposition:
\begin{eqnarray}
\label{totFT}
\bF=\sum_{i=1}^N \bF_i\:, \qquad \bT=\sum_{i=1}^N \left({\bT}_i+ \mathbf{r}_i\times\bF_i\right)\:,
\end{eqnarray}
where $\mathbf{r}_i$ is the radius vector to the centre of $i$th particle in a chosen coordinate frame.


\end{document}